\documentclass[12pt]{article}
\usepackage{epsfig}
\usepackage{amssymb}
\usepackage{epsfig}

\begin{document}
\def\be{\begin{equation}}
\def\ee{\end{equation}}
\def\ba{\begin{eqnarray}} 
\def\ea{\end{eqnarray}}
\def\nn{\nonumber}

\newcommand{\bbf}{\mathbf}
\newcommand{\rrm}{\mathrm}

\title{ A variational approach to the low energy properties of even-legged d-dimensional quantum spin 
systems\\} 
\author{Jean Richert
\footnote{E-mail address: richert@lpt1.u-strasbg.fr}\\ 
Laboratoire de Physique Th\'eorique, UMR 7085 CNRS/ULP,\\
Universit\'e Louis Pasteur, 67084 Strasbourg Cedex,\\ 
France} 
\date{\today}
\maketitle 
\begin{abstract}

We develop a variational formalism in order to study the structure of low energy spectra of frustrated 
quantum spin systems. It is first applied to trial wavefunctions of ladders with one spin $1/2$
on each site. We determine energy minima of these states. The variational ground state shows a finite 
energy gap with respect to the energies of states which span the Hilbert space and are orthogonal to 
it. This is the case for any size of the system. Under some justifiable approximations the 
argument can be extended to even-legged ladder systems in $2d$ and higher dimensional spaces. 
The Hamiltonian can contain spin-spin coupling interactions of any range. For specific values of the 
coupling strengths level degeneracies can occur.

\end{abstract} 
\maketitle
%
%
\\

\section{Introduction.}

The low energy spectra of magnetic materials can be described by quantum spin models characterized 
by their geometric structure, the form of the interactions between the spins located on sites, the
strength and the range of the couplings which fix the intensity of the interactions. The knowledge 
of the spectral properties of these systems is considered to be of prime importance for the 
understanding of specific phenomena such as superconductivity at high $T_c$. Investigations on this 
subject which go on for many years have concentrated in the recent past on specific structures, in 
particular the presence of ladders and stripes in $2d$ superconducting material, see f.i.
~\cite{hot,tas,hay,tran}.\\  

It has been observed that systems which expectedly behave like 2-leg ladders show a gap between
the ground state and the first excited state ~\cite{dag} whereas those which behave like 3-leg 
ladders possess a continuous spectrum~\cite{azu}. Theoretical investigations have been developed in 
order to study the properties of the ground state and low energy states with a particular interest 
for the existence or absence of such a gap. In $1d$ systems Lieb, Schultz and Mattis (LSM) 
~\cite{lieb} established that spin-$1/2$ systems which are invariant under translations and rotations 
either show no finite energy gap between the ground state and the first excited state, or the ground 
state is degenerate and spontaneouly breaks the translational invariance creating a dimerized 
structure. In the first case the energy interval between the ground state and the first excited state 
$\Delta E$ goes to zero as $1/L$ when the length $L$ of the chain tends to $\infty$. A large amount 
of work on spectral properties of quantum spin systems followed and can be found in the litterature 
from the late 60s to the late 80s, see f. i. ~\cite{ham,schu,aff,and,tim}.\\
 
An important step was performed by Haldane who conjectured that the spectra of Heisenberg 
antiferromagnetic chains are gapless if the spins are half-integers and show a gap if the spins are 
integers~\cite{hald1,hald2}. Later Shelton et al.~\cite{she} found that weakly coupled isotropic 
spin-$1/2$ antiferromagnetic Heisenberg chains always show a spectral gap. In an application of the 
non-linear sigma-model Sierra~\cite{sie} tried to extend the Haldane conjecture to ladders. He showed 
that in the framework of this model spin-$1/2$ even ladders exhibit a gapped 
spectrum whereas odd ladders show a gapless spectrum. Further tests of Haldane's conjecture have been 
made, see f.i.~\cite{cab1, cab2} and refs. therein. Oshikawa extended the LSM theorem to quantum 
spin systems with a conserved particle number on a periodic lattice in arbitrary dimensions 
and showed that an excitation gap opens when the number of particles per unit cell of the 
ground state is an integer~\cite{oshi}. The theorem was also found to be verified in 
chiral liquid spin systems in higher dimensions~\cite{sore}. There it was however observed that 
the conclusions drawn in ref.~\cite{oshi} may not be universally verified. More recently 
Hastings~\cite{hast} showed that $\Delta E$ goes to zero like $lnL/L$ for spin-$1/2$ systems in 
higher dimensions when the Hamiltonian of the system is $SU(2)$  invariant and sites are occupied 
by an odd number of spin-$1/2$ particles.\\ 

Further information has been obtained by means of specific models, in particular quantum spin
systems, ladders and 2$d$ networks, see in particular ~\cite{kak,ner,kol,bat,sta}. These studies deal 
with specific systems using different formal tools. They show non trivial low energy spectral 
properties and very rich phase diagrams. However the low energy properties of $2d$ and higher
quantum spin systems remain an open question in a large number of cases.\\
   
In the present work we propose a further investigation concerning the existence of a finite energy 
gap $\Delta E$ between the ground state and the low energy states of frustrated quantum spin systems. 
More specifically we consider spin $1/2$ ladders and extend the analysis to even-legged $2d$ systems,
as well as systems with an even number of chains in an arbitrary number of space dimensions. We show 
that the results are valid for any range of the spin-spin interaction.\\ 

The investigation relies on a variational approach. We consider first the case of spin-$1/2$  
ladders. The central concept which governs the behaviour of the system is SO(4)-invariance of the 
Hamiltonian. We introduce a trial wavefunction in order to determine the energy extremum of the system 
and analyze the nature of this extremum. If this state corresponds to a minimum in energy we show that 
the states orthogonal to it are characterized by an energy gap with respect to this state
except for specific values of the coupling constants which enter the Hamiltonian leading to degeneracies.
We then show how the results can be extended to $2d$ and higher dimensional spaces for systems 
with an even number of legs. Finally we comment and discuss the outcome of the present investigations 
and draw conclusions.\\ 
 
\section{The model.}
 
We consider spin-$1/2$ ladders~\cite{lin1,lin2} described by Hamiltonians of the following type 
\ba\label{eq0}
H^{(s,s)} &=& J_t\sum_{i=1}^{N} s_{i_1}s_{i_2} + J_l\sum_{<ij>}s_{i_1}s_{j_1} +
J_l\sum_{<ij>}s_{i_2}s_{j_2} + J_{1c}\sum_{<ij>}s_{i_1}s_{j_2}  
\\ \nn
&  & + J_{2c}\sum_{<ij>}s_{i_2}s_{j_1}
\ea
where the indices $1$ or $2$ label the spin $1/2$ vector operators $s_{i_k}$ acting on the sites $i$ on 
both ends of a rung, in the second and third term $i$ and $j$ label nearest neighbours along the legs 
of the ladder. The fourth and fifth term correspond to diagonal interactions between nearest sites 
located on different legs. $N$ is the number of sites on a ladder and the coupling strengths 
$J_t, J_l, J_{1c}, J_{2c}$ are positive. In the sequel we fix $J_{1c} = J_{2c} = J_{c}$.
The ladder is represented in Fig. 1.\\

The Hamiltonian (1) possesses $SO(4)$ symmetry by construction. By means of a spin rotation 
~\cite{kika}
\be
s_{i_1} = \frac{1}{2} (S_i + R_i)
\label{eq1} \ .
\ee
\be
s_{i_2} = \frac{1}{2} (S_i - R_i)
\label{eq2} \ .
\ee
it can be expressed in the form
\be
H^{(S,R)} = \frac{J_t}{4}\sum_{i=1}^{N} (S_{i}^{2} -  R_{i}^{2}) + J_1\sum_{<ij>}S_i S_j + 
J_2\sum_{<ij>}R_i R_j
\label{eq3} \   
\ee
where the components $S_{i}^{(+)}, S_{i}^{(-)}, S_{i}^{(z)}$ and $R_{{i}}^{(+)}, R_{{i}}^{(-)},
R_{{i}}^{(z)} $ of the vector operators $S_{i}$ and  $R_{i}$ are the $SO(4)$ group generators
which can be written as

\ba\nonumber
S_{i}^{(+)}  = \sqrt{2}(X^{(11)(10)}_{i} + X^{(10)(1-1)}_{i}) = S_{i}^{(-)*}
\ea
\ba\nonumber
S_{i}^{(z)}  =  X^{(11)(11)}_{i} - X^{(1-1)(1-1)}_{i}
\ea
\ba\nonumber
R_{i}^{(+)}  =  \sqrt{2}(X^{(11)(00)}_{i} + X^{(00)(1-1)}_{i}) = R_{i}^{(-)*}
\ea
\ba\nonumber
R_{i}^{(z)} =  - (X^{(10)(00)}_{i} + X^{(00)(10)}_{i})
\ea 

where 
\ba\nonumber
X^{(SM_{s})(S' M_{S'})}_{i} = |SM_{s}\rangle_{i}._{i}\langle S' M_{S'}|\nonumber
\ea
and the states $|SM_{S}\rangle$ are generated by the coupling of spin $1/2$ states   
\ba\nonumber
|SM_{s}\rangle_{i} = \sum_{m_1,m_2} \langle 1/2 ~~m_1 ~~1/2 ~~m_2| 1 ~0\rangle
|1/2 ~~m_1\rangle_{i}  |1/2 ~~m_2\rangle_{i}
\ea
to two particle singlet ($S = 0, M_{S} =0$) and triplet ($S = 1, M_{S} = 0, +1, -1$)
states. 
\\ 

\begin{figure}
\epsfig{file=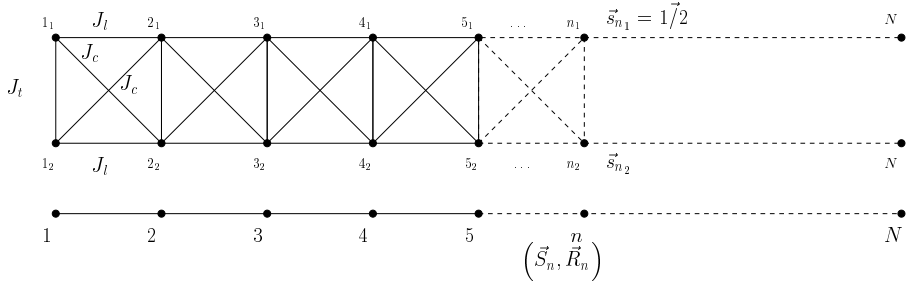,width=12cm}
\caption{Top: the original spin ladder. The coupling strengths are indicated as given in 
the text. Bottom: The ladder in the SO(4) representation. See the text.}
\end{figure}

In the new representation the coupling stengths are $J_1 = (J_l + J_{c})/2$ and  $J_2 = (J_l - J_{c})/2$.
$J_1$ is always positive, the sign of $J_2$ depends on the relative strength of the couplings  
between spins on sites located on the same leg ($J_l$) and different legs ($J_{c}$). The $SO(4)$
symmetry is shown to be of particular interest for the characterization of the spectral properties of
even-legged spin systems. This point appears clearly below.\\

\section{Variational ansatz.}

\subsection{N-body wavefunction.}

Variational techniques have been applied for a long time in order to investigate the properties of 
quantum spin systems, see f. i. refs.~\cite{and,gro1,hee,gro2,yok1,yok2,bec,pam,xian}. 
In analogy with the phenomenological BCS approach and the approach of refs.~\cite{maj1,maj2,kol,anle} 
we introduce a trial wavefunction which takes a product form of two spin-$1/2$ states
\be 
|\Phi_{0}^{(N)}\rangle = \prod _{i=1}^{N} |\Psi^{(0)}_i\rangle
\label{eq4} \ .
\ee
Here $|\Psi^{(0)}_i\rangle$ is the wavefunction at the location $i$ which is chosen as a linear
combination of singlet and triplet states $| S M_s \rangle_{i}$
\be 
|\Psi^{(0)}_i\rangle = \alpha_i| 0 0\rangle _i+  \beta_{i{\overline1}}| 1 -1\rangle_i + 
\beta_{i0}| 1 0\rangle_i + \beta_{i 1}| 1 +1\rangle_i
\label{eq5} \ .
\ee
with the normalization
\be  
|\alpha_i|^2 + |\beta_{i{\overline1}}|^2 + |\beta_{i 0}|^2+ |\beta_{i 1}|^2 = 1
\label{eq6} \ .
\ee

The time-reversal symmetry of the Hamiltonian allows to introduce 
$|\beta_{i{\overline1}}|^2= |\beta_{i 1}|^2$. A further constraint is imposed such that 
$ \beta_{i 1} = \beta_{i{\overline1}}=\kappa \beta_{i 0}$  where $\kappa$ is fixed as 
the ratio 
\be  
\kappa = \frac{\langle 1/2 ~~1/2 ~~1/2 ~~1/2| 1 ~~1\rangle}  
{\langle 1/2 ~~1/2 ~~1/2 ~-1/2| 1 ~0\rangle} = \sqrt{2}
\label{eq7} \ .
\ee
where the Clebsh-Gordan coefficients weigh the coupling of the two electron state 
$ |1/2 m_1 ; 1/2 m_2\rangle_i$ to the states $|1 1\rangle_i$ and $|1 0\rangle_i$
respectively. Each site is occupied by a single spin. The constraint takes care of the relative 
weight of the $| 1 ~1\rangle$ and the $| 1 ~0\rangle$ two-spin states. Defining 
$\gamma_{i} = (1 + 2 \kappa^{2})^{1/2} \beta_{i 0}$ leads to the normalization relation  
\be  
|\alpha_i|^2 +  |\gamma_i|^2 = 1 
\label{eq8} \ .
\ee
\\

The imposed constraint fixes $\beta_{i 1} = \kappa\gamma_{i}/(1 + 2 \kappa^{2})^{1/2} = 
\beta_{i{\overline1}}$. The $\{\alpha_i\}$ and $\{\gamma_i\}$ are the variational quantities.\\

Using the expressions of the operators $S_i$ and $R_i$ defined above the energy 
$E_{N}^{0} = \langle\Phi_{0}^{(N)}|H^{(S,R)}|\Phi_{0}^{(N)}\rangle$ associate to the trial 
wavefunction $|\Phi_{0}^{(N)}\rangle$ is obtained as 
\be  
E_{N}^{0} =  2 E_T \sum_{i=1}^{N}|\gamma_{i}|^2 + E_S \sum_{i=1}^{N} (3 |\alpha_i|^2 + |\gamma_{i}|^2)
+ E_{int1} +  E_{int2}
\label{eq9}\ .
\ee
with $E_T = J_t/4$, $E_S = - J_t/4$,
\be  
E_{int1}=  \frac{8 \kappa^2 J_1}{(1 + 2 \kappa^{2})^2}\sum_{<ij>}|\gamma_{i}|^2 |\gamma_{j}|^2
\label{eq10} \ .
\ee 
and
\ba\label{eq11} 
E_{int2} & = & \frac{ 4 J_2}{(1 + 2 \kappa^2)}\sum_{<ij>} (Re(\alpha_i \gamma_i^{*}) 
Re(\alpha_j \gamma_j^{*}))   
\\ \nn
&  & + \frac{ 8 \kappa^2   J_2 }{ (1 + 2 \kappa^2)} \sum_{<ij>} (Im(\alpha_i^{*} \gamma_i) 
Im(\alpha_j \gamma_j^{*}))
\ea
\\

\subsection{Extrema of the energy}

In order to fix the energy spectrum of the system the complex amplitudes $\alpha_i$ and $\gamma_i$
are varied under the normalization constraint Eq.~(\ref{eq8}) 
\be  
(\frac{d}{d\alpha_i^{*}} + \frac{d\gamma_i}{d\alpha_i^{*}}\frac{d}{d\gamma_i}) E_{N}^{0} = 0 
\label{eq12} \ .
\ee 
\\
We introduce the parametrization
\be  
\alpha_i = exp(i\phi_{1i}) sin(\theta_i/2)  
\label{eq13} \ .
\ee 
and
\be  
\gamma_i  = exp(i\phi_{2i}) cos(\theta_i/2)  
\label{eq14} \ .
\ee
Applying Eq.~(\ref{eq12}) to $E_N^{0}$ given by Eq.~(\ref{eq9}) and using Eq.~(\ref{eq13})
and Eq.~(\ref{eq14}) leads to 
\be  
\epsilon_{i} exp(i(\phi_{1i} - \phi_{2i}) sin(\theta_i) + (cos^2(\theta_i/2) -
sin^2(\theta_i/2)) \Delta_i = 0
\label{eq15} \ .
\ee
where  
\be  
\epsilon_i =(E_S - E_T) - \frac{4 \kappa^2 J_1}{(1 + 2 \kappa^2)^2} \sum_{<j>} 
|\gamma_j|^2
\label{eq16} \ .
\ee
The energy $\epsilon_{i}$ is a negative quantity, $\Delta_i$ is complex and given by
\be  
\Delta_i =  \frac{J_2}{1 + 2 \kappa^2} \sum_{<j>} ((1 - 2 \kappa^2)  \alpha_j ^{*}\gamma_j 
+ (1 + 2 \kappa^2 )\alpha_j \gamma_j ^{*}) 
\label{eq17} \ .
\ee
The symbol $<j>$ indicates the nearest neighbour of $i$, in practice $j = i+1$.
$\Delta_i$ can be written as
\be  
\Delta_i = |\Delta_i| exp(i(\phi_{1i} - \phi_{2i}))
\label{eq18} \ .
\ee
The solution of Eq.~(\ref{eq15}) takes the form
\be  
\epsilon_{i} sin \theta_i  + |\Delta_i| cos \theta_i  = 0
\label{eq19} \ 
\ee
with 
\be  
sin \theta_i  = - |\Delta_i|/ E_i
\label{eq20} \ 
\ee
and
\be  
\cos \theta_i  = \epsilon_i/ E_i. 
\label{eq21} \ 
\ee
Here $E_i^2 = \epsilon_i^2 + |\Delta_i|^2$ and\\
\be  
|\gamma_i|^2 = \frac{1}{2} (1 + \frac{\epsilon_i }{E_i})
\label{eq 22} \ .
\ee
\be  
|\alpha_i|^2 = \frac{1}{2} (1 - \frac{\epsilon_i }{E_i})
\label{eq 23} \ .
\ee
\be  
\Delta_i = \Delta_i^{r} + i \Delta_i^{i}
\label{eq24} \ .
\ee
\be  
\Delta_i^{r} = -\frac{ J_2}{(1 + 2 \kappa^2)}\sum_{<j>} \frac{\Delta_j^{r}}{E_j }
\label{eq25} \ .
\ee
\be  
\Delta_i^{i} = -\frac{2 \kappa^{2} J_2}{(1 + 2 \kappa^2)}\sum_{<j>} \frac{\Delta_j^{i}}{E_j }
\label{eq26} \ .
\ee
\\

The energy $E_i$ can in principle be positive or 
negative. The expression of $\epsilon_i$ in Eq.~(\ref{eq16}) shows that the variational 
wavefunction $|\Phi_{0}^{(N)}\rangle$ postulated in Eq.~(\ref{eq4}) entangles nearest neighbouring 
site state wavefunctions.\\

The Hamiltonian of the spin ladder is fixed by the three coupling constants $J_t$, $J_1$ and $J_2$.
A close inspection of the expressions of $\epsilon_i$ (Eq.~(\ref{eq16})), $\Delta_i$
(Eq.~(\ref{eq17})) shows that the amplitude $\alpha_i$ and hence $\gamma_i$ of each site
can be determined self-consistently if the system is closed in such a way that site $N+1$ is identified 
with site $1$.\\

\subsection{Nature of the extrema.}

The nature of the extremum is given by the sign of the second derivative $d^2 E_{N}^{0}$
of the energy $ E_{N}^{0}$ with respect to $\alpha_i$ under the constraint fixed by Eq.~(\ref{eq8}) 
and the definition of $\gamma_i$. The explicit calculation leads to
\\
\be  
 d^2 E_{N}^{0} = \frac{J_t}{4} (-3 + \frac{|\alpha_i|^2}{|\gamma_i|^2})+  
\frac{|\alpha_i|^2}{|\gamma_i|^2} 
\frac{8 \kappa^2 J_1}{(1 + 2 \kappa^2)^2}\sum_{<j>}|\gamma_j|^2 
\label{eq27} \ .
\ee 
\\
The sign of $d^2 E_{N}^{0}$ depends on the relative strengths of $J_t$, $J_1$ 
which are positive quantities and the ratio of the amplitudes $|\alpha_i|^2$ and $|\gamma_i|^2$. 
In fact
\be
\frac{|\alpha_i|^2}{|\gamma_i|^2} = \frac{E_i - \epsilon_i}{E_i + \epsilon_i}
\label{eq28}\ .
\ee 
which  is larger or smaller than $1$ depending on the sign of $E_i$. 
\\ 
 
$d^2 E_{N}^{0}$ is positive if $J_t$ is small enough so that the first term 
in Eq.~(\ref{eq27}) gets smaller than the second term which is strictly positive.\\ 

More precisely, for fixed $J_t$ and $J_1$ $d^2 E_{N}^{0}$ is positive if
 $|\alpha_i|^2/|\gamma_i|^2$ gets larger than a minimum value,
\\
\be\nonumber
|\alpha_i|^2/|\gamma_i|^2 \geq \frac{3} {1 + \frac{4}{J_{t}} F(\kappa, J_{1})}
\ee
where
\\
\ba\nonumber
F(\kappa, J_{1}) = \frac{8 \kappa^2 J_1}{(1 + 2 \kappa^2)^2}\sum_{<j>} |\gamma_j|^2
\ea
\\
This may be realized if $J_1$ is large and (or) $|\Delta_i|$ is small enough which may be the case 
for small $|J_2| = |(J_l - J_{c})|/2$.\\
 
\section{Energy of the variational state. Nature of the spectrum.}
 
\subsection{Unprojected wavefunction.}

Using the explicit expressions of the amplitudes  $\{\alpha_i\}$ and $\{\gamma_i\}$  leads to 
the expression of the variational energy

\ba\label{eq29}
E_{N}^{0} &=& (E_T + 2E_S)N + (E_T - E_S)\sum_{i}\frac{\epsilon_i}{E_i}
\\ \nn
&  & + \frac{2 \kappa^2 J_1}{(1 + 2 \kappa^2)^2}\sum_{i} (1 + \frac{\epsilon_i }
{E_i}) \sum_{<j>} (1 + \frac{\epsilon_j }{E_j})
\\ \nn
&  & + \frac{ J_2}{(1 + 2 \kappa^2)}\sum_{i} \frac{\Delta_i^{r}}{E_i}
\sum_{<j>} \frac{\Delta_j^{r}}{E_j} - \frac{2 J_2 \kappa^2}{(1 + 2 \kappa^2)}
\sum_{i} \frac{\Delta_i^{i}}{E_i} \sum_{<j>} \frac{\Delta_j^{i}}{E_j}
\ea
 
We want to analyze the properties of the energy spectrum, in particular to see whether it is 
discrete or continuous. \\

Since the Hamiltonian of the system is real symmetric the space of states is spanned by a 
complete set of states which are orthogonal to each other.  

\ba\nonumber
|\Phi_{k}^{(N)}\rangle = \prod _{i=1}^{N} |\Psi^{(k)}_i\rangle 
\ea

Orthogonality is realized if $|\Phi_{k}^{(N)}\rangle$ differs from $|\Phi_{0}^{(N)}\rangle$ 
by at least one state

\be 
|\Psi_i^{(k)}\rangle = \alpha_i^{(k)}| 0 0\rangle _i + \gamma_i^{(k)}| 1 0\rangle_i  
\label{eq30} \ .
\ee

orthogonal to 
\be 
|\Psi_i^{(0)}\rangle = \alpha_i^{(0)}| 0 0\rangle _i + \gamma_i^{(0)}| 1 0\rangle_i  
\label{eq31} \ .
\ee 
where $\alpha_i^{(0)}$ and  $\gamma_i^{(0)}$ stands for the former notation $\alpha_i$ and 
$\gamma_i$ of section $3$.\\

If $\alpha_i^{(k)} = - \gamma_i^{(0)*}$  and  $\gamma_i^{(k)} = \alpha_i^{(0)*}$ the wavefunction 
in which $|\Psi_i^{(0)}\rangle$ is replaced by $|\Psi_i^{(k)}\rangle$
is then orthogonal to it by construction. Using $|\Psi_i^{(k)}\rangle$ the energy 
$\epsilon_{(i-1)}^{(k)}$ corresponding to the location of site $(i-1)$  of the state 
$|\Phi_{k}^{(N)}\rangle$ orthogonal to the corresponding state $|\Psi_i^{(0)}\rangle$
in $|\Phi_{0}^{(N)}\rangle$ is then given by ($\sum_{<j>}$ corresponds in fact to the 
site $i$, see below Eq.~(\ref{eq17})
\be  
\epsilon_{(i-1)}^{(k)} = (E_S - E_T) - \frac{4 \kappa^2 J_1}{(1 + 2 \kappa^2)^2} 
|\gamma_{i}^{(k)}|^{2}
\label{eq 32} \ .
\ee
where $|\gamma_{i}^{(k)}|^{2} = |\alpha_{i}^{(0)}|^{2}$. Eq. (34) shows that
in general
\be  
\epsilon_{(i-1)}^{(k)} \not=\epsilon_{(i-1)}^{(0)}
\label{eq 33} \ .
\ee
Using Eq.~(\ref{eq17}) it is easy to see that 
\be  
\Delta_{i-1}^{(k)} = - \Delta_{i-1}^{(0)}
\label{eq 34} \ .
\ee
and hence
\be  
E_{(i-1)}^{(k)} \not=E_{(i-1)}^{(0)}
\label{eq 35} \ .
\ee
\\
 
Consequently the states $|\Psi_i^{(0)}\rangle$ and $|\Psi_i^{(k)}\rangle$  are 
non degenerate states since 
\ba\label{eq36}
E_{N}^{k} - E_{N}^{0}&=& (E_T - E_S)(\frac{\epsilon_{(i-1)}^{(k)}}{E_{(i-1)}^{(k)}}
- \frac{\epsilon_{(i-1)}^{(0)}}{E_{(i-1)}^{(0)}})
\\ \nn
&  & + \frac{2 \kappa^2 J_1}{(1 + 2 \kappa^2)^2}( 1 + \frac{\epsilon_i}{E_i})
(\frac{\epsilon_{(i-1)}^{(k)}}{E_{(i-1)}^{(k)}} - \frac{\epsilon_{(i-1)}^{(0)}}
{E_{(i-1)}^{(0)}})
\\ \nn
&  & + \frac{ J_2}{(1 + 2 \kappa^2)}\frac{\Delta_{i}^{r}}{E_i}
(\frac{\Delta_{i-1}^{r(k)}}{E_{(i-1)}^{(k)}} - \frac{\Delta_{i-1}^{r(0)}}{E_{(i-1)}^{(0)}})
\\ \nn 
&  & - \frac{2 \kappa^{2} J_2}{(1 + 2 \kappa^2)}\frac{\Delta_{i}^{i}}{E_i}
(\frac{\Delta_{i-1}^{i(k)}}{E_{(i-1)}^{(k)}} - \frac{ \Delta_{i-1}^{i(0)}}{E_{(i-1)}^{(0)}})
\ea
\\
is different from $0$ whatever the number of sites except if $|\Delta_{i-1}^{(k)}|  = 
|\Delta|_{i-1}^{(0)}|  = 0$ which may happen if $J_2 = 0$  i. e. $J_l = J_{c}$.\\ 

The physical states which are orthogonal to the variational state are generally linear 
combinations of the orthogonal states $|\Phi_k^{(N)}\rangle$. 
Since these states are non-degenerate with the variational ground state this property 
will be shared by the physical states in the general case, i.e. there is a finite gap 
between the variational state and the other physical states. As already mentioned above 
degeneracy in energy may occur in special cases, f.i. if $J_2= 0$.\\ 

\subsection{Projected wavefunction.}

Since the Hamiltonian $H^{(S,R)}$ commutes with the total spin projection $M_{tot}$ the 
actual wavefunction has to be projected on a state of fixed total spin projection
\be 
{\cal P} |\Phi_{0}^{(N)}\rangle = \frac{1} {2\pi} \int _{-\pi}^{\pi} {d\phi e^{i \phi(M_{tot} - 
\sum _{i} S_{i}^{z})}} \prod _{i = 1}^{N} \sum _{k = 1} ^{4} \eta _{k}^{i}
| S _{k}  M _{k} \rangle_{i}
\label{eq37} \ .
\ee

where $\eta_{1}^{i}=\alpha_{i}, \eta_{2}^{i}=\beta_{i{\overline1}},\eta_{3}^{i}=\beta_{i0},
\eta_{4}^{i}=\beta_{i1}$.\\

The energy corresponding to the projected state ${\cal P} |\Phi_{0}^{(N)}\rangle$ is then 
obtained from the hermitic matrix element

\be 
E_{P}^{0} = \frac{1}{2} [\langle\Phi_{0}^{(N)}|H^{(S,R)}{\cal P}|\Phi_{0}^{(N)}\rangle
+ \langle\Phi_{0}^{(N)}|{\cal P} H^{(S,R)}|\Phi_{0}^{(N)}\rangle] 
\label{eq38} \ .
\ee

In the case where $M_{tot} = 0$ one gets 
 
\be 
\langle\Phi_{0}^{(N)}|H^{(S,R)}{\cal P}|\Phi_{0}^{(N)}\rangle = \sum_{p=0}^{\infty} 
\frac{(-1)^{p}}{(2p)!!} \frac{\pi^{2p}}{2p+1} \sum _{l_{1},...,l_{n}}
c^{l}_{l_{1}...l_{n}}{\cal E}_{N}^{0} (l_{1},..,l_{n})
\label{eq39} \ .
\ee

where $c^{l}_{l_{1}...l_{n}}$ is the combinatorial coefficient associated with the
decomposition of $\sum_{i} (S^{z}_{i})^{l_i}$, $l = 2p = \sum_{i} l_{i}$ and \\

\be
{\cal E}_{N}^{0} (l_{1},..,l_{n}) = \langle\Phi_{0}^{(N)}|H^{(S,R)} \prod_{i = 1}^{N} 
\sum _{k = 1} ^{4}\eta _{k}^{i} (M_{ki})^{l_{i}}| S _{k}  M _{k} \rangle_{i}
\label{eq40} \ .
\ee
where $\{M_{ki}\}$ are the projections of the states $k$ on the sites $i$.  

Using the explicit expressions of the $\{\alpha_i\}$, $\{\gamma_i\}$ and 
$\{\Delta_i\}$ (eqs. (23) to (27)) ${\cal E}_{N}^{0} (l_{1},..,l_{n})$ can be put in the form

\ba\label{eq41}
& &{\cal E}_{N}^{0} (l_{1},..,l_{n}) = \frac{(2 E_T + E_S)}{ \lambda^2}\sum_{i}
[\delta_{l_{i}0} + \kappa^2(1 + (-1)^{l_i}] |\gamma_i|^2       
\\ \nn
&  &+ 3 E_S \sum_{i}  \delta_{l_{i}0}|\alpha_i|^2
+ \frac{\kappa^2 J_1}{\lambda^4} \sum_{<ij>} A(i,j) |\gamma_i|^2 |\gamma_j|^2
\\ \nn
& + & \frac{J_2}{\lambda^2} \sum_{<ij>}[C(i,j) \alpha_{i}^{*}\alpha_{j}^{*}
\gamma_{i}\gamma_{j} - \delta_{l_{i}0}\delta_{l_{j}0}\alpha_{i} \alpha_{j}
\gamma_{i}^{*}\gamma_{j}^{*} + B(j,i)\alpha_{i} \alpha_{j}^{*}\gamma_{i}^{*}
\gamma_{j} + B(i,j)\alpha_{i}^{*} \alpha_{j}\gamma_{i}\gamma_{j}^{*}]
\ea
\\

where $\lambda = (1 + 2 \kappa^{2})^{1/2}$ and
\ba\label{eq42}  
A(i,j) &=& \kappa^2 [1 + (-1)^{l_i+l_j} + (-1)^{l_i+1} + (-1)^{l_j+1}] 
\\ \nn
&  & + 2\delta_{l_{i}0}\delta_{l_{j}0} + \delta_{l_{i}0}(1 + (-1)^{l_i} + 2(-1)^{l_i})
\ea
 
\be  
B(i,j)= \delta_{l_{j}0} [1 + (-1)^{l_i} + \delta_{l_{i}0}]
\label{eq43} \ . 
\ee
and
\be  
C(i,j)= \delta_{l_{i}0}\delta_{l_{j}0} + (-1)^{l_i+1} + (-1)^{l_j+1}
\label{eq44} \ .
\ee
\\

One can then construct projected states $\{{\cal P} |\Phi_{k}^{(N)}\rangle\}$ which are 
orthogonal to ${\cal P} |\Phi_{k0}^{(N)}\rangle$ in the same way as in section $4a)$ above
and find out that the energy difference $\Delta{\cal E} = {\cal E}_{k}^{N} - {\cal E}_{0}^{N}$
and the corresponding $\Delta E_{P} = E_{P}^{k} - E_{P}^{0}$ are different from zero
except maybe if $J_2 = 0$.\\

\section{Generalization to even-legged spin 1/2 systems in 2d with any type (short or long) range 
interaction.} 

It is straightforward to see that the present considerations can be applied to spin ladders which 
do not only contain interactions between neighbouring sites, but also interactions at any distance 
from each other. Such contributions introduce a quantitative but no qualitative change in 
the gap equations Eqs.~(\ref{eq24}), (\ref{eq25}) and (\ref{eq26}).
 
More generally, the results can be extended to $2d$ frustrated systems with an even number of 
ladder legs showing the same type of couplings between sites as those introduced on the ladder. 
They are obtained by adding explicit couplings between sites located on the neighbouring legs which 
belong to neighbouring ladders. If $L= M*N$ where $M$ is the number of parallel ladders and
$N$ the length of the ladders, the Hamiltonian in the spin rotated representation 
(Eqs. (2) and (3)) can be put in the form\\  
 
\be
H^{(S,R)} = \frac{J_t}{4}\sum_{i=1}^{L} (S_{i}^{2} -  R_{i}^{2})+ J_1\sum_{<ij>}S_i S_j + 
J_2\sum_{<ij>}R_i R_j + H^{(int)}
\label{eq45} \ .
\ee 

with $H^{(int)} = H^{(int1)} + H^{(int2)}$ and

\be
H^{(int1)}=\frac{J_{int}^{r}}{4}\sum_{i=1}^{L-N} (S_i S_{i+N} + S_i R_{i+N} - R_i S_{i+N} 
- R_i R_{i+N})
\label{eq46} \ .
\ee 
\ba\label{eq47}
H^{(int2)} &=&  \frac{J_{int}^{d}}{4} \left[\sum_{i=1}^{L-N-1}\left(S_i S_{i+1+N} + 
S_{i} R_{i+N+1}   \right. \right.\\ \nn
& & \left. \left.- R_{i} S_{i+N+1} - R_i  R_{i+1+N}\right )\right.\\ \nn
& & \left. + \sum_{i=2}^{L-N}\left (S_{i-1} S_{i+N} + S_{i-1} R_{i+N} - R_{i-1} S_{i+N} 
- R_{i-1} R_{i+N}\right)\right]
\ea
\\ 
$J_{int}^{r}$ is a coupling strength between sites in neighbouring chains 
belonging to neighbouring ladders along rungs and $J_{int}^{d}$ acts between the same elements 
along a diagonal linking neighbouring site (see Fig. 2).\\ 
 
\begin{figure}
\epsfig{file=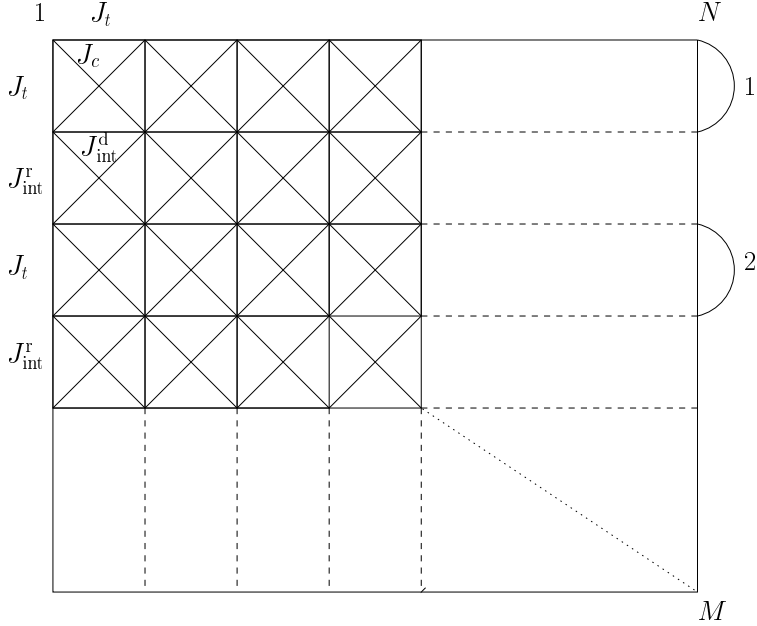,width=10cm}
\caption{The even-legged 2d spin system. The coupling strengths between successive ladders are 
indicated by $J_{int}^{d}$ and $J_{int}^{r}$ as written in the text.}
\end{figure}

The Hamiltonian Eq.~(\ref{eq45}) to Eq.~(\ref{eq47}) describes a unique chain
which is constructed in such a way that the end of each chain $m$ $(m = 1,...M)$
corresponding to the Hamiltonian $ H^{(S,R)}$  Eq.~(\ref{eq3}) is connected to the beginning of
the next one, see Fig. (3).\\ 

\begin{figure}
\epsfig{file=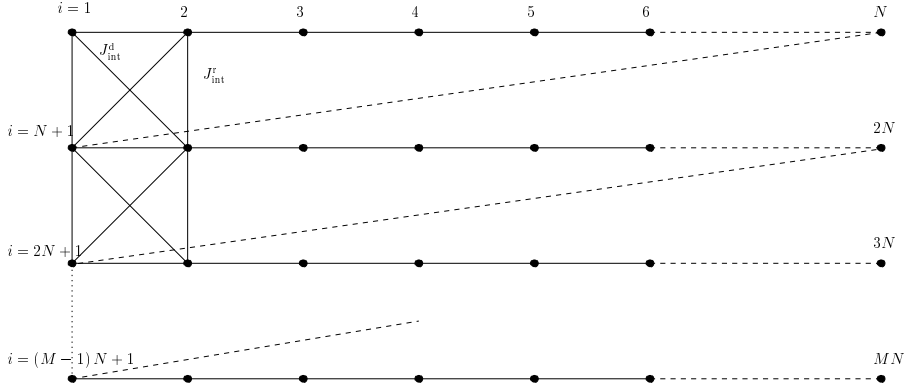,width=12cm}
\caption{The 2d system of fig. 2 in the SO(4) representation. The long dashed links are the 
spurious contributions. See explanations and discussion in the text.}
\end{figure}

This introduces spurious terms $S_{kN} S_{(k+1)N}$, $R_{kN} R_{(k+1)N}$,
$R_{kN} S_{(k+1)N}$, $S_{kN} R_{(k+1)N}$, $k = 1,...,M-1$ which couple the border sites of
the consecutive spin chains. The contributions of their matrix elements should be subtracted 
from the expression of the energy after the variation. The number of terms of this type is 
however small compared to the total number of states if $M$ and $N$ are large ("surface" 
contributions). Neglecting them in the limit $L = M*N \rightarrow \infty$ one can go through the 
calculations as before, obtain the same type of equations and draw the same type of conclusions 
as before.\\

More explicitly the expression of the energies is given by 
\ba\label{eq48}   
\epsilon_i & = &(E_S - E_T) - \frac{4 \kappa^2 J_1}{(1 + 2 \kappa^2)^2} \sum_{<j>} 
|\gamma_j|^2
\\ \nn
&  & - \frac{ \kappa^2 J_{int}^{r}}{(1 + 2 \kappa^2)^2}\sum_{(j)}|\gamma_j|^2
- \frac{ 2 \kappa^2 J_{int}^{d}}{(1 + 2 \kappa^2)^2}\sum_{[j]}|\gamma_j|^2
\ea
 and the complex gap $\Delta_i$ which is now the sum of three terms

\be   
\Delta_i = \Delta_{1i} + \Delta_{2i} + \Delta_{3i}
\label{eq49} \ .
\ee
where 
\be
\Delta_{1i} = - \frac{J_2}{(1 + 2 \kappa^2)} \sum_{<j>} ((1 - 2 \kappa^2)\alpha_j ^{*}\gamma_j  
+(1 + 2 \kappa^2)\alpha_j \gamma_j ^{*}) 
\label{eq50} \ .
\ee
\be
\Delta_{2i} = - \frac{J_{int}^{r}}{4(1 + 2 \kappa^2)} \sum_{(j)}((1 - 2 \kappa^2)
\alpha_j ^{*}\gamma_j + (1 + 2 \kappa^2)\alpha_j \gamma_j ^{*})
\label{eq51} \ .
\ee
\be
\Delta_{3i} = - \frac{J_{int}^{d}}{2(1 + 2 \kappa^2)}\sum_{[j]}((1 - 2 \kappa^2)
\alpha_j ^{*}\gamma_j + (1 + 2 \kappa^2)\alpha_j \gamma_j ^{*})
\label{eq52} \ .
\ee
$<j>$ ($= i + 1$) stands for the site which is the right-nearest neighbour along a chain ,
$(j)$ corresponds to the nearest-neighbour of a site $i$ along a rung between chains belonging 
to nearest-neighbour ladders and $[j]$ to the cross-diagonal nearest-neighbour of site $i$ 
located on the chain of the nearest-neighbour ladder.\\
 
The expression of the variational energy is formally the same as the one of section $4$ except 
for the additional sum of terms corresponding to the couplings $J_{int}^{r}$ and $J_{int}^{d}$. 
It is given by
\ba\nonumber
d^2 E_{N} &=& \frac{J_t}{4} (-3 + \frac{|\alpha_i|^2}{|\gamma_i|^2})+  
\frac{|\alpha_i|^2}{|\gamma_i|^2} 
\frac{2 \kappa^2}{(1 + 2 \kappa^2)^2} [4 J_1 \sum_{<j>}|\gamma_j|^2
\\ \nn
& & + J_{int}^{r}\sum_{(j)}|\gamma_j|^2 + 2 J_{int}^{d}\sum_{[j]} |\gamma_j|^2] 
\ea
\\
The nature of the extremum (maximum or minimum) depends on $J_t$, $J_1$ and the additional 
coupling strengths $J^{r}_{int}$ and  $J^{d}_{int}$. $d^2 E_{N}$ is positive if  
$|J_{2}| = |J_{l} - J_{c}|/2$, $J_{int}^{r}$ and $J_{int}^{d}$ are such that 
\ba\nonumber 
|\alpha_i|^2/|\gamma_i|^2 \geq \frac{3} {1 + \frac{4}{J_{t}} 
G(\kappa, J_{1}, J_{int}^{r}, J_{int}^{d})}
\ea
where
\be 
G(\kappa, J_{1}, J_{int}^{r}, J_{int}^{d}) = 
\frac{2 \kappa^2 }{(1 + 2 \kappa^2)^2} [4 J_1 \sum_{<j>}|\gamma_j|^2
+ J_{int}^{r}\sum_{(j)}|\gamma_j|^2 + 2 J_{int}^{d}\sum_{[j]} |\gamma_j|^2] 
\label{eq53} \ 
\ee
\\
 
Hence $d^2 E_{N}$ is positive if $J_{l}$, $J_{int}^{r}$, $J_{int}^{d}$ get large and $J_{t}$ and 
(or) the $\{\Delta_i\}$s remain relatively small.\\

The discussion about the structure of the spectrum can be taken from section 4. In general this 
spectrum will show a gap between the ground state and excited states, except for specific values of 
specific coupling strengths. In fact degeneracy of states appears when the gaps $\Delta_{ni}$ 
($n = 1,2,3$) disappear, which happens when $J_2$, $J^{r}_{int}$ and $J^{d}_{int}$ are equal to 
zero.\\ 
 
The present results can evidently be extended to the case where one works with projected wavefunctions  
along the same lines as it has been shown in section 4.\\

\section{Remarks, discussion.}

 The present derivations are subject to comments:
\begin{itemize}
  
\item The spin Hamiltonians $H$ which were introduced commute with also with the total 
spin $S_{tot}$. 
Hence the corresponding quantum numbers are conserved and characterize the states of the physical 
spectrum  $|\Phi_{0}^{(N)}(S_{tot}, M_{tot})\rangle$. The trial wavefunction  $|\Phi_{0}^{(N)}\rangle$ 
should be projected on a state with fixed $S_{tot}$. This changes quantitatively the energy $E_N^{0}$  
and $E_{P}^{0}$. It should however not qualitatively affect the calculated energy differences and the 
existence of a gap between the variational ground state and other states of the infinite system.\\     
 
\item In practice  $\beta_{i 1}=\beta_{i{\overline1}}$ may be treated as independent variational 
parameters if the constraint introduced in section $3a)$  Eq.~(\ref{eq7}) is not taken into account.
The corresponding component may be kept in the definition of $|\Psi_{i}^{(0)} \rangle$,
Eq.~(\ref{eq5}).\\  

\item It is easy to see that the present formalism can be extended to any even-legged system 
of spins in any space dimension $D = nd$ with the proviso that, as in the $2d$ case, the contribution
of terms coming from the border sites of the finite systems and which should not be present
can be neglected when applying the variational procedure. This is realized if the system gets large. 
Since the algebraic expressions are rather cumbersome we do not write them out explicitly.  

\item The "linking procedure" used in Section 5 in order to generate a continuous linear system
leads a posteriori to a symmetry constraint of $2d$ and systems of higher dimensionality which by 
construction possess the $SO(4)$ symmetry introduced in section 2. This is a constraint which may 
limit the range of the variational procedure in the sense that other symmetries could take
over at different space dimensionalities.\\ 

\end{itemize}

\section{Degeneracies.}

\subsection{Behaviour of $\epsilon_i$ and $\Delta_i$ at $J_2 = 0$.}

As observed above in sections  $4$ and $5$ the variational state gets degenerate with states 
orthogonal to it when the gaps $\{\Delta_i\}$ vanish. This is the case when $J_2 = 0$ (section $4$), 
$J_2 = J_{int}^{d} = J_{int}^{nd} = 0$ (section $5$). It is of interest to analyze more 
precisely the property of this point. In the sequel we concentrate on the case of a 
ladder system. The generalization to $2d$ and higher dimensions is straightforward.\\   

It is easy to determine the behaviour of $\Delta_i$ and $\epsilon_i$ in the neighbourhood of 
$J_2 = 0$. The derivative of each gap with respect to $J_2$ reads
\ba\label{eq54}
\frac{d \Delta_i}{dJ_2} &=& - \frac{1}{1 + 2 \kappa^2}\frac{\Delta_{i+1}(J_2)}{E_{i+1}(J_2)} 
- \frac{J_2}{1 + 2 \kappa^2} ( \frac{\Delta_i^{'}(J_2)}{E_{i+1}(J_2)} 
\\ \nn
&  & - \Delta_{i+1}(J_2) \frac{\epsilon_{i+1}(J_2) \epsilon_{i+1}^{'}(J_2)
+ \Delta_{i+1}(J_2)\Delta_{i+1}^{'}(J_2)}{E_{i+1}^{3}(J_2)})
\ea
where $\Delta_i^{'}(J_2)= d\Delta_i(J_2)/dJ_2$ and $\epsilon_i^{'}(J_2) = d\epsilon_i(J_2)/dJ_2$.
$\Delta_i^{'}(J_2) = 0$ for $J_2 = 0$ and is a continuous function of $J_2$ in the 
neighbourhood of this point.\\ 

Similarly $\epsilon_i^{'}(J_2)$ is a continuous function of 
$J_2$ in the neighbourhood of this point.\\  


\subsection{Eigenstates at $J_2 = 0$.} 

We show that the states $\{|\Phi_{k}^{(N)}\rangle\}$, $k = 0$ and $k \neq 0$ are eigenstates of the 
Hamiltonian $H^{(S,R)}$ defined in Eq.(4) when $J_2 = 0$.\\

Indeed the part of the Hamiltonian proportional to $J_t$ is diagonal in the basis of states 
$\{|\Phi_{k}^{(N)}\rangle\}$, the third term does not contribute and the second term 
is a sum of terms of the form 
\be
\langle\Phi_{k}^{(N)}|S_i S_j|\Phi_{k'}^{(N)}\rangle = \frac{8 \kappa^{2}}{\lambda^4}
\gamma_{i}^{(k)*}\gamma_{j}^{(k)*}\gamma_{i}^{(k')} \gamma_{j}^{(k')}
\label{eq55} \ .
\ee 

Using the expression of $\gamma_i$ in Eq. (23) and the negative root solution 
of $E_i$, $E_i = - (\epsilon_i^2 + |\Delta_i|^2)^{1/2}$ leads to 
$\langle\Phi_{k}^{(N)}|S_i S_j|\Phi_{k'}^{(N)}\rangle = 0$ for any $k$ and  $k'$, $i$ and $j$.
Hence the states $\{|\Phi_{k}^{(N)}\rangle\}$ are eigenstates of $H^{(S,R)}$ orthogonal to each 
other.\\




\subsection{Behaviour of the energy interval between the states in the neighbourhood of 
$J_2 = 0$.}
 
The energy intervals between  the physical states which are linear combinations of states
$\{|\Phi_{k}^{(N)}\rangle\}$ vary continuously with $J_2$ in the neighbourhood of 
$J_2 = 0$.\\ 

Using the expression given in  Eq.~(\ref{eq36}) for a state $|\Phi_{k}^{(N)}\rangle$
which differs from the state $|\Phi_{k'}^{(N)}\rangle$ by $n$ consecutive site 
wavefunctions $|\Psi^{(k)}_{l}\rangle$, (f.i. $l = N-n+1,...,N$) the energy differences 
between these states are given by

\ba\label{eq56}
\Delta E_{N}^{(kk')} &=&  E_{N}^{(k)} - E_{N}^{(k')} = (E_T - E_S)\sum_{l}(e_{l}^{(k)} - e_{l}^{(k')})
\\ \nn
&  & + \frac{2 \kappa^2 J_1}{(1 + 2 \kappa^2)^2} \sum_{l}[(1 + e_{N-l})(e_{l}^{(k)} - e_{l}^{(k')})
\\ \nn
&  & + (e_{l}^{(k)} + e_{l+1}^{(k)} -(e_{l}^{(k')} + e_{l+1}^{(k')}) +  
e_{l}^{(k)} e_{l+1}^{(k)} -  e_{l}^{(k')}e_{l+1}^{(k')}]
\\ \nn
&  & + \frac{ J_2}{(1 + 2 \kappa^2)} [\delta_{N-n}^{r} (\delta_{N-n+1}^{r(k)} -
\delta_{N-n+1}^{r(k')}) +  \sum_{l} \delta_{l}^{r(k)} (\delta_{l+1}^{r(k)} -  
\delta_{l+1}^{r(k')})]
\\ \nn
&  &
- \frac{2 \kappa^{2} J_2}{(1 + 2 \kappa^2)}[\delta_{N-n}^{i} (\delta_{N-n+1}^{i(k)} -
\delta_{N-n+1}^{i(k')}) +  \sum_{l} \delta_{l}^{i(k)} (\delta_{l+1}^{i(k)} -  
\delta_{l+1}^{i(k')})]
\ea 
where $e_{l}^{(k)} = \epsilon_{l}^{(k)}/E_{l}^{(k)}$, $\delta_{l}^{r(k)} = 
\Delta_{l}^{r(k)}/E_{l}^{(k)}$ and $\delta_{l}^{i(k)} = \Delta_{l}^{i(k)}/E_{l}^{(k)}$
depend implicitly on $J_2$.\\

$\Delta E_{N}^{(kk')}$ goes continuously to zero for any $\{k,k'\}$ when $J_2$ goes to 
zero which shows that the degeneracy of states for this particular value of the coupling 
$J_2$. The same is true for the physical states which go continuously over to the states 
$\{|\Phi_{k}^{(N)}\rangle\}$ when  $J_2$ goes to zero. This is also the case when the 
states $|\Psi^{(k)}_{l}\rangle$ which differ in different states $|\Phi_{k'}^{(N)}\rangle$ 
are non consecutive as considered above.\\

Close to $J_2 = 0$  $E_{N}^{(k)}$  and hence $\Delta E_{N}^{(kk')}$ behave like
$cnst * J_{2}^{2}$ to lowest order in  $J_2$ and $E_{N}^{(k)}(J_2)\simeq E_{N}^{(k)}(-J_2)$ \\        

Finally the same reasoning is valid in the projected framework dealing with 
${\cal P} |\Phi_{k}^{(N)}\rangle$ and ${\cal P} |\Phi_{k'}^{(N)}\rangle$.\\

\section{Generalization to a correlated variational wavefunction.} 

\subsection{Correlated wavefunction.}
 
The variational wavefunction which was postulated in Eq.~(\ref{eq4}) has a mean-field structure
which includes implicit correlations between site states through Eq.~(\ref{eq19}) to 
Eq.~(\ref{eq26}) but no explicit correlations. In order to implement such correlations we  
go over to a pseudo-fermion representation of the operators $S_{i}^{(+)}, S_{i}^{(-)}, S_{i}^{(z)}$
and $R_{{i}}^{(+)}, R_{{i}}^{(-)}, R_{{i}}^{(z)}$.

\ba\nonumber
S_{i}^{(+)} = b_{i}^{+} d_{i} + d_{i}^{+} c_{i}
\ea 
\ba\nonumber
S_{i}^{(-)} = c_{i}^{+} d_{i} + d_{i}^{+} b_{i}
\ea 
\ba\nonumber
S_{i}^{(z)} = b_{i}^{+} b_{i} + c_{i}^{+} c_{i} + d_{i}^{+} d_{i}
\ea
\ba\nonumber
R_{i}^{(+)} = b_{i}^{+} a_{i} + a_{i}^{+} c_{i}
\ea
\ba\nonumber
R_{i}^{(-)} = a_{i}^{+} b_{i} + c_{i}^{+} a_{i}
\ea 
\ba\nonumber
R_{i}^{(z)} = c_{i}^{+} a_{i} + a_{i}^{+} d_{i} 
\ea 
where $a_{i}^{+}, b_{i}^{+}, c_{i}^{+}, d_{i}^{+}$ are anticommuting fermion operators which 
generate singlet and triplet states

\ba\nonumber
a_{i}^{+}|0\rangle_{i} = |S = 0, M = 0\rangle_{i}
\ea 
\ba\nonumber
b_{i}^{+}|0\rangle_{i} = |S = 1, M = 1\rangle_{i}
\ea 
\ba\nonumber
c_{i}^{+}|0\rangle_{i} = |S = 1, M = -1\rangle_{i}
\ea 
\ba\nonumber
d_{i}^{+}|0\rangle_{i} = |S = 1, M = 0\rangle_{i}
\ea 
 
and $|0\rangle_i$ is the particle vacuum on site $i$.\\

We introduce the variational trial wavefunction
\be
|\Upsilon_{0}^{(N)}\rangle = e^{F}|0\rangle
\label{eq57} .\
\ee  
where 
\be
F = f_{(1)}\sum_{i_1}\Omega_{i_1}^{+} + f_{(2)} 
\sum_{i_1}\sum_{i_2}\Omega_{i_1}^{+}\Omega_{i_2}^{+} + f_{(3)} 
\sum_{i_1}\sum_{i_2}\sum_{i_3}\Omega_{i_1}^{+}\Omega_{i_2}^{+}\Omega_{i_3}^{+}
+....
\label{eq58} .\ 
\ee 
$N$ is the total number of sites in the $SO(4)$ representation, $|0\rangle$ the 
particle vacuum and

\ba\nonumber
\Omega_{i_l}^{+} = \alpha_{i_l}a_{i_l}^{+} + \beta_{i_l}b_{i_l}^{+} +
\gamma_{i_l}c_{i_l}^{+} + \delta_{i_l}d_{i_l}^{+} 
\ea 
\ba\nonumber
[\Omega_{i_l}^{+}, \Omega_{i_m}^{+}] = 0
\ea
\\

Developing the exponential in Eq.~(\ref{eq57}) leads to 

\be
|\Upsilon_{0}^{(N)}\rangle =  [1 + F + \frac{1}{2!}F^{2} +  \frac{1}{3!}F^{3} + ...]
|0\rangle
\label{eq59} .\ 
\ee 

Since the creation operators $\{\Omega_{i}^{+}\}$ are fermions and since the system should be 
occupied by two particles by site $i$ (1 particle on each site of a rung $i$) the sum of terms 
in Eq.~(\ref{eq58}) and  Eq.~(\ref{eq59}) is restricted to a finite number of terms which all 
contain exactly $N$ distinct operators $\{\Omega_{i}^{+}\}$. This sum is written in a fixed order 
so that $i_1<i_2<i_3<...<i_N$. The state $|\Upsilon_{0}^{(N)}\rangle$ can be written as 
\be
|\Upsilon_{0}^{(N)}\rangle = \sum_{p=1}^{N} \frac{1}{p!} \sum^{'}_{n_1,n_2,...n_N}  
B_{N}(p;n_1,n_2,...,n_N)f_{(1)}f_{(2)}...f_{(N)} \Omega_{1}^{+}...\Omega_{N}^{+} 
|0\rangle
\label{eq60} .\ 
\ee 
where the sum contains only the terms which generate $N$ operators $\{\Omega_{i}^{+}\}$
and
\ba\nonumber
B_{N}(p;n_1,n_2,...,n_N) = (-1)^{P} \frac{N!}{n_{1}!...n_{p}!} 
\ea
with the constraint $n_1+2n_2+....+ Nn_N=N$ and $P$ the number of permutations such that the 
operators $\{\Omega_{i}^{+}\}$ appear in increasing order with respect to their index,
$i_1<i_2<i_3<...<i_N$.\\
 
In a realistic description  the correlation functions $f_{(l)}$ could depend on the relative 
distance between the particles on the different sites, $f_{(l)} = f^{(i_1,...i_l)}_{(l)}$.
We do not take this fact into account. It is sensible to believe that this point is not crucial 
for the aim which we fix here, i.e. to show the qualitative behaviour of the system as already 
claimed before. It makes also physically sense to consider that the coefficients $f_{(l)}$, 
($l \geq 2$) which correlate $l$ particles are a product of two-body terms 
$f_{(l)} = C(l,2) f_{(2)}$ where $C(l,2) = l!/(2!(l-2)!)$.\\ 

Under these conditions

\be
|\Upsilon_{0}^{(N)}\rangle = \sum_{p=1}^{N}\frac{1}{p!}\sum^{'}_{n_1,n_2,...,n_N}  
B_{N}(p;n_1,...,n_N)f_{(1)}^{n_1}f_{(2)}^{n_{(2..N)}}
\prod_{l=2}^{N} [C(l,2)]^{n_l}|\Phi_{0}^{(N)}\rangle 
\label{eq61} .\ 
\ee 

where $n_{(2..N)} = n_2+...+n_N$ with the constraint $n_1+2n_2+....Nn_N=N$.\\
 
\subsection{Variational wavefunction.}

The structure of the correlated wavefunction $|\Upsilon_{0}^{(N)}\rangle$ is a product of 
a term which contains the correlations and the mean-field wavefunction $|\Phi_{0}^{(N)}\rangle$. 
In its whole generality the search of extrema of the energy 
$E_{N}^{(c)} = \langle\Upsilon_{0}^{(N)}|H^{(S,R)}|\Upsilon_{0}^{(N)}\rangle$ is in principle 
given by varying the coefficients $\{\alpha_i\}$, $\{\beta_i\}$, $\{\gamma_i\}$, $\{\delta_i\}$
and $f_{(1)},f_{(2)}$ under the normalization constraint
\be
\langle\Upsilon_{0}^{(N)}|\Upsilon_{0}^{(N)}\rangle = 1 
\label{eq62} .\ 
\ee
\\
 
One may consider a two-step procedure suggested by the structure of the wavefunction 
$|\Upsilon_{0}^{(N)}\rangle$ in Eq. (63). A minimization on the wavefunction $|\Phi_{0}^{(N)}\rangle$
is done first on $E_{N}^{(0)} = \langle\Phi_{0}^{(N)}|H^{(S,R)}|\Phi_{0}^{(N)}\rangle$ 
as before in section 4. Then Eq.~(\ref{eq62}) is verified if 
\ba\label{eq63}
\cal C &=&  \sum_{p,p'}\frac{1}{p!p^{'}!} \sum_{{n_i},{n'_i}}
\prod_{l=2}^{N}[ C(l,2)]^{2n_l}
B_{N}(p;n_1,...,n_N) B_{N}(p';n^{'}_1,...,n^{'}_N) 
\\ \nn
& & f_{(1)}^{n_1 + n^{'}_{1}}f_{(2)}^{n_{(2..N)} + n^{'}_{(2..N)}} = 1
\ea
where the sums over $\{n_i\}$ are such that $n_1+2n_2+....+ Nn_N=N$ and the same for
$\{n^{'}_i\}$.\\ 

$\cal C$ should then be fixed by looking for values of $f_{(1)}$ and $f_{(2)}$ such that the 
normalization constraint given by  Eq.~(\ref{eq62}) is realized and $E_{N}^{(c)}$ is minimized.\\

By construction it appears that all the reasoning concerning the nature of the spectrum which was 
made in sections $4$ and $5$ remains then valid as long as solutions exist. Hence the same conclusions
as those drawn there can be taken up in the present case.\\

If such a solution can be found it ensures the existence of a gap for any space dimension.\\
 
\section{Summary and conclusions.}

We introduced a variational approach in order to describe the low-lying energy states of frustrated
quantum spin systems. We started with spin-$1/2$ ladders and extended the approach to even-legged 
systems in any space dimension $d$.\\
 
In a first step the postulated trial wavefunction was written as a superposition of spin singlet and 
triplet states in the framework of $SO(4)$ symmetry. The variational ansatz limited the wavefunction to 
a mean-field product of site states formally similar to the ansatz used in high-$T_c$ theory~\cite{pwa}. 
The local amplitudes of the site state components were correlated through nearest neighbour contributions. 
We looked for energy extrema and analyzed the conditions under which the variational procedure leads 
to energy minima. The realization of this property depends on the parameter subspace in which the 
coupling strengths of the spin-spin interactions are located. If the energy extrema do not correspond 
to minima the variational ansatz used for the trial wavefunction does not describe the structure of 
the physical ground state. In these cases the structure of this state must be different from the 
postulated ansatz.\\

In the cases where the trial wavefunction describes a state of minimum energy the present analysis shows 
that spectra of spin-$1/2$ quantum spin systems with two legs are gapped. The approach can in principle 
be generalized to spin-spin interactions of any range and any type of frustration if the postulated 
variational ground state wavefunction leads to an energy minimum. For specific values of the coupling 
constants the gap may vanish leading to degeneracies in energy.\\  

The extension of the formalism to even-legged systems of dimension $d \geq 2$ was developed in section 5.
The pertinence of such an extension which has been discussed in section 6 is warranted as long as
the $SO(4)$ symmetry which characterizes even-legged ladders remains a symmetry which governs the system
at higher space dimensions.\\ 

In a further step we suggested an extension of the variational trial wavefunction to include explicit  
correlations beyond the mean-field approximation. We introduced simplifications to the most general
form of this wavefunction which may have quantitative but not qualitative effects on the conclusions.
It is seen that the corresponding spectra may again show the characteristic gap obtained with the 
mean-field variational ansatz.\\  

The author would like to thank Daniel Cabra and Konstantin Kikoin for comments and critical remarks.

\end{document}